# Periodogram and likelihood periodicity search in the SNO solar neutrino data

Gioacchino Ranucci and Marco Rovere
*Istituto Nazionale di Fisica Nucleare*
*Via Celoria 16, 20133 Milano, Italy*
e-mail: gioacchino.ranucci@mi.infn.it, marco.rovere@mi.infn.it

ABSTRACT

In this work a detailed spectral analysis for periodicity search of the time series of the $^8$B solar neutrino flux released by the SNO Collaboration is presented. The data have been publicly released with truncation of the event times to the unit of day (1 day binning); they are thus suited to undergo the traditional Lomb-Scargle analysis for periodicity investigation, as well as an extension of such a method based on a likelihood approach. The results of the analysis presented here confirm the absence of modulation signatures in the SNO data. For completeness, a more refined "1 day binned" likelihood is also illustrated, which approximates the unbinned likelihood methodology, based upon the availability of the full time information, adopted by the SNO collaboration. Finally, this work is completed with two different joint analyses of the SNO and Super-Kamiokande data, respectively, over the common and the entire data taking periods. While both analyses reinforce the case of the constancy of the neutrino flux, the latter in addition provides evidence of the detection at the 99.7% confidence level of the annual modulation spectral line due to the Earth's orbit eccentricity around the Sun

# I. INTRODUCTION

Recently the SNO collaboration published the results of the periodicity search performed on the measured time series of the $^8$B neutrino flux [1], and afterwards released publicly the raw data used in the analysis [2]. In this work we present the results obtained performing alternative periodicities investigations on the released datasets. The basic tool adopted for this investigation is the estimation of the frequency spectrum of the neutrino flux times series, since it is well known that periodicities hidden in the data would appear as sharp peaks in the spectrum itself. The important and crucial part of the analysis, however, is not the calculation of the spectrum, but the precise assessment of the statistical significance of the most prominent experimental spectral peaks, since the noise affecting the data series produces erratic peaks which can attain also very high values.

Hence a large part of the paper is devoted to such significance evaluation, performed following a Monte Carlo procedure well established in [3] and [9]. Such an analysis is complemented by the determination of the sensitivity of the method for the discovery of a true oscillation embedded in the data, which will obviously depend upon the amplitude and frequency of the hypothetical signal.

The paper is organized as follows: in paragraph II a brief description of the data is given (essentially a short summary of the thorough description reported in [1]), in paragraph III there is the illustration of the two different data analysis procedures adopted, paragraph IV is devoted to the significance assessment of the largest detected peaks via the Monte Carlo determination of the null hypothesis distributions, in paragraph V the sensitivity of the methods for the discovery of true modulations is reported, in paragraph VI the features of the analysis results at two specific frequencies of interest are described, in paragraph VII it is illustrated a procedure which approximates the characteristics of an unbinned likelihood analysis, in paragraph VIII it is reported a joint analysis of the SNO and Super-Kamiokande I data over the common data taking period, in paragraph IX the joint analysis of the two experiments is extended over the entire, respective data taking periods, and finally in paragraph X there are the conclusions.

# II. DESCRIPTION OF THE DATA SETS

It is well known that the SNO detector is an heavy water Cerenkov detector which measures $^8$B solar neutrinos via charged-current and neutral-current interactions on deuterons in 1 kton of $D_2O$, and also through neutrino-electron elastic scattering interactions.

The datasets used in this analysis, officially released by the SNO collaboration, are similar to those used in the periodicity search reported in [1]. They are related to the two first phases of the experiment, i.e. the pure $D_2O$ phase and the salt phase, when salt has been added to the heavy water in order to increase the neutron detection efficiency. The data taking period started on November 2, 1999 and encompassed in total about 1400 calendar days.

Specifically the SNO collaboration released for both phases two files, one with the number of daily detected events, and the other with the specification of the runtime periods during each day of data taking. A proper periodicity analysis clearly has to take into account the actual live time within each day of data taking in order to properly normalize the detected counts. It must be pointed out that there is a difference between the released data and those used by the SNO Collaboration in its periodicity search: while the latter retain the full time information, the former are characterized by the rounding of the time of each event to the units of day. Practically this implies that the data have been published in the format of 1 day binning.

The event rates vs. time, obtained trough a suitable bin manipulation similar to that used in [1], are displayed in Fig. 1 and 2 for the $D_2O$ and salt phase, respectively. In particular, some of the nominal 1 day bins have been merged together when the respective width was much shorter than the nominal value, in order to avoid large scattering in the data display. The rate is 9.35 ev/day in the case of the $D_2O$ phase, and 11.85 ev/day in the case of the salt phase. The fact the shape of the two figures reproduces well also the details of the two corresponding figures in [1] gives confidence that the retrieval and arrangement of the data have been done properly.

In particular, each point of the series in the two figures is characterized by a time value which is the weighted mean time of the corresponding bin, evaluated so to take into account properly the bin live time, and by a corresponding ordinate value which is the normalized count rate in the same bin, i.e. the total counts in that bin divided by the corresponding live time.

This summary description of the data is enough for the purpose of the present analysis. For further details on the characteristics of the datasets the reader can refer to [1].

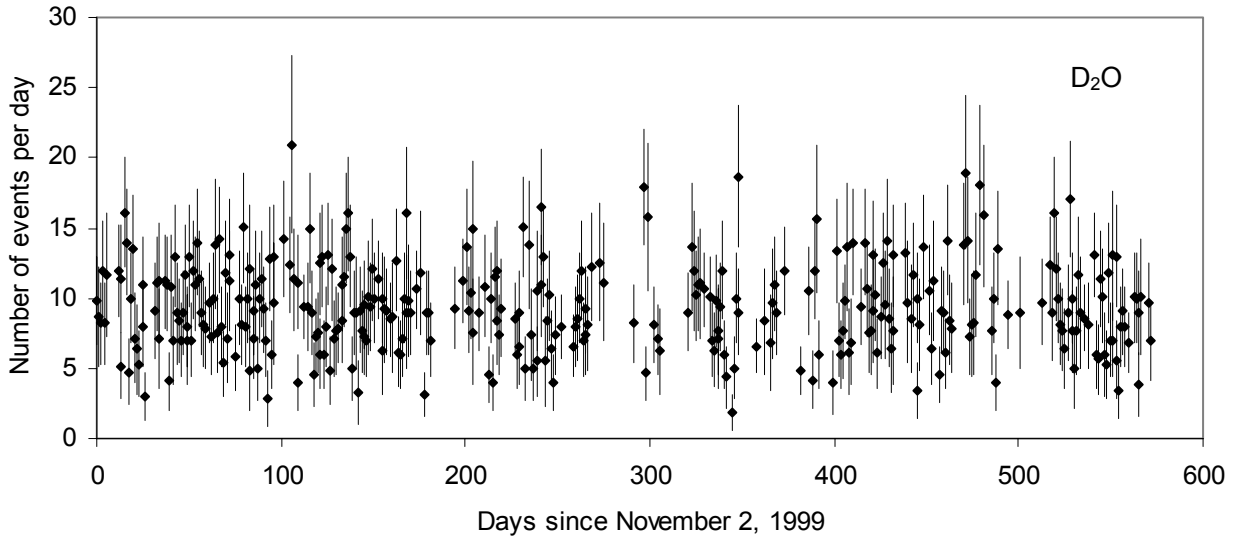

*Fig. 1 – $D_2O$ phase event rate. The mean rate is 9.35 ± 0.17 ev/day.*

### III. SPECTRAL DATA ANALYSIS METHODS AND SPECTRA

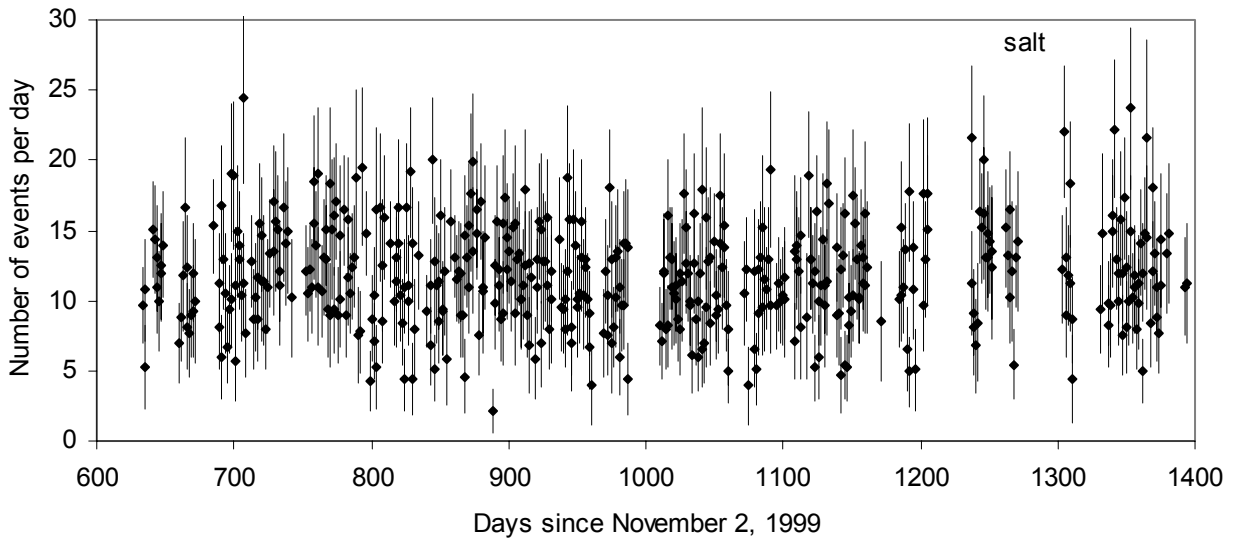

*Fig. 2 – Salt phase event rate. The mean rate is 11.85 ± 0.17 ev/day.*

In the periodicity search published by SNO the main tool of analysis is an unbinned maximum likelihood scan of the data, which exploits fully the time information for each neutrino event. Because of the rounding in the released data of the neutrino event times to the unit of day mentioned above, the same exact analysis methodology cannot be applied here. Furthermore, in [1] a second analysis has been presented based on the so called weighted Lomb-Scargle method [4].

Instead, for the present analysis two alternative approaches have been chosen, which somehow complement those adopted by the SNO collaboration: a) the standard (not weighted) Lomb-Scargle periodogram search and b) a likelihood extension of the Lomb-Scargle search itself.

### A. Lomb-Scargle Method

#### 1. Statistical properties

Traditionally the Lomb-Scargle spectrum (denoted more frequently as periodogram) is the standard tool for the spectral analysis of time series composed by unevenly spaced points.

The data as displayed in Fig. 1 and 2 are already arranged in a suitable way to undergo such a standard methodology; indeed, the data in this format reproduce an unevenly sampled time series, thanks to the arrangement to represent an extended bin via a single weighted mean live time (it may be worth to remind that the same format was used in the data released by the Super-Kamiokande collaboration [5]). On the other hand, the ordinate values may be considered as sampled by a continuous function representing the count rate vs. time.

By denoting with $t_r$ the sequence of time values, with $x_r$ the corresponding measured values and with $F$ the average value of the whole series, the Lomb-Scargle periodogram is computed as [6] [7] [9]

$$\frac{1}{2\sigma^2}\left(\frac{\left(\sum_{r=1}^{N}(x_r - F)\cos\omega(t_r - \tau)\right)^2}{\sum_{r=1}^{N}\cos^2\omega(t_r - \tau)} + \frac{\left(\sum_{r=1}^{N}(x_r - F)\sin\omega(t_r - \tau)\right)^2}{\sum_{r=1}^{N}\sin^2\omega(t_r - \tau)}\right) \quad (1)$$

where

$$\frac{\sum_{r=1}^{N}\sin 2\omega t_r}{\sum_{r=1}^{N}\cos 2\omega t_r} = \tan 2\omega\tau \quad (2)$$

and

$$\sigma^2 = \frac{1}{N-1}\sum_{r=1}^{N}(x_r - F)^2 \quad (3).$$

The Lomb-Scargle periodogram is characterized by well defined statistical properties described for example in [7] and recently reviewed in [8]. Here the main aspects of such properties are reminded. In case of the null hypothesis that the series is sampled by a pure Gaussian noise process with no modulation embedded, the periodogram ordinate at each frequency follows a simple exponential distribution $e^{-h}$; such an exponential distribution is the basis of the false alarm probability formula introduced in [7] and [9] to perform the significance assessment of the largest detected peak in the periodogram. This false alarm formula, which is

$$P(>z) = 1 - (1 - e^{-z})^M \tag{4}$$

where *M* is the number of independent scanned frequencies in the search for periodicities, stems from the probability density function of the height of the largest spectral peak in case of pure noise series: if the frequencies are *M*, each individually exponentially distributed, the PDF (probability density function) of the largest value is obviously

$$p_{largest}(h) = M\left[1 - e^{-h}\right]^{M-1} e^{-h} \tag{5}$$

from which the integration above a threshold z gives eq. (4).

Hence in practice, the quantification of the consistency (or inconsistency) of a time series with the hypothesis of constant rate according to the Lomb-Scargle method requires simply to compute its spectrum, to note the maximum spectral peak $z_{max}$ and to evaluate, by substituting $z_{max}$ in eq. (4), the probability to get an ordinate as high or higher than $z_{max}$ by pure noise chance: as a consequence the statistical significance of the peak is high (i.e. likely it is actually due to a true signal) if the false alarm formula gives a low or very low value.

This methodology is not ambiguous if *M* is known. However *M* is exactly known only in the particular case of even sampling [9]. In the practical cases of uneven sampling *M* is not easily a-priori defined: it depends on the coherence of the series and it is heuristically interpreted as the effective number of independent scanned frequencies; in such a situation the above simple analytical procedure is more effectively replaced by a Monte Carlo approach (see for example [3] and [9]): many modulation free synthetic time series with the same timing and statistical (in this case of Poisson nature) properties of the experimental series under study are generated, for each of them the relevant spectrum is computed, and the maximum values of all the spectra are used to create the histogram which represents the so called null hypothesis distribution, i.e. the maximum spectral peak distribution under the assumption of pure noisy series. The significance of the largest spectral peak $z_{max}$ in the actual spectrum of the experimental series is then obtained reading off the Monte Carlo histogram the fraction of times in which the simulation produced a value as high or higher than $z_{max}$ itself.

The Monte Carlo output can also be used to determine the number *M* of effectively scanned frequencies, being for this purpose simply required to evaluate the value of *M* which ensures the best fit of the simulated null hypothesis distribution to the relation (5).

The standard approach described above considers only the highest spectral peak as diagnostic element to judge the consistency (or inconsistency) of the series with the constant rate assumption. More insight on the feature of the series, however, can be obtained by extending such a statistical treatment also to the other spectral peaks (practically, to the bunch of the more prominent peaks). In particular, by ordering the peaks at the various frequencies over the search band in term of their height, it is possible to generalize eq. (5) in order to express analytically the distribution of the height of the lowest peak, of the second lowest peak and so on, up to the highest peak, in case of pure white noise time series. How to perform such generalization (in the different context of photoelectron statistics ) has been shown in [10] and [11]: still denoting the number of independent frequencies with *M*, the probability density function of the spectrum ordinate of the $i_{th}$ (in term of height, starting from the lowest) spectral peak is

$$p_i(h/M) = \frac{M!}{(i-1)!(M-i)!}\left[1 - F(h)\right]^{(M-i)}\left[F(h)\right]^{i-1} p(h) \tag{6}$$

where

$$F(h) = \int_0^h p(\lambda)d\lambda \qquad (7)$$

being *p(h)* simply $e^{-h}$.

As cross check it can be easily verified that eq. (6) reduces to (5) in case of the highest peak.

The Monte Carlo extension of the procedure is quite straightforward, since it requires only to evaluate, in addition to the null hypothesis distribution of the highest peak, those of the second highest, third highest and so on peaks (in the present analysis such an extension has been limited to the first four highest peaks). The model functions (6) in the following will be used to model the curves obtained from the Monte Carlo calculations.

It should be noted that, strictly speaking, the statistical properties of the Lomb-Scargle periodogram summarized in this section are valid in the assumption of time series affected by Gaussian noise; instead, the count process generating the time series of the measured SNO data is obviously of Poisson nature. The comparison of the model with the Monte Carlo in paragraph 4 will show, however, that the above results maintain their validity also in this case.

## 2. Experimental spectra

Due to the 1 day binning of the released data, the corresponding frequency analysis is characterized by a Nyquist frequency which is about 0.5 cycle/days (or in term of years about 180 cycles/years). The spectra presented in this section (as well as those related to the likelihood methodology) are thus limited up to such frequency boundary (note: the ordinate of the spectra is conventionally denoted as Power).

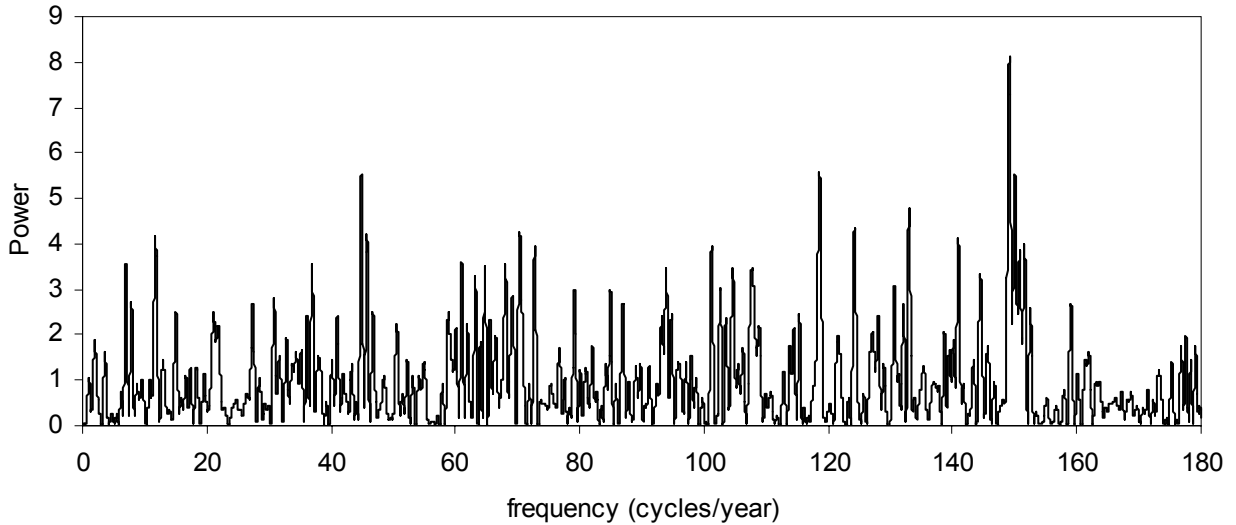

*Fig. 3 – $D_2O$ phase Lomb-Scargle periodogram.*

The three periodograms related to the $D_2O$ phase, the salt phase and the combined dataset obtained via application of the relations (1), (2) and (3) are shown in Fig. 3, 4 and 5 (the analysis of the combined dataset has been performed after scaling the values of the salt period in order to equalize the average count rate of both phases).

In the spectrum in Fig. 3 the four highest peaks have, respectively, ordinates 8.11 (149.13), 5.59 (118.6), 5.55 (44.86) and 5.54 (150.03) (in the brackets there are the corresponding frequencies expressed in cycles/year).

In the spectrum in Fig. 4 the major peaks are, respectively, 7.43 (156.52), 6.16 (28.15), 5.55 (0.75) and 5.32 (52.93).

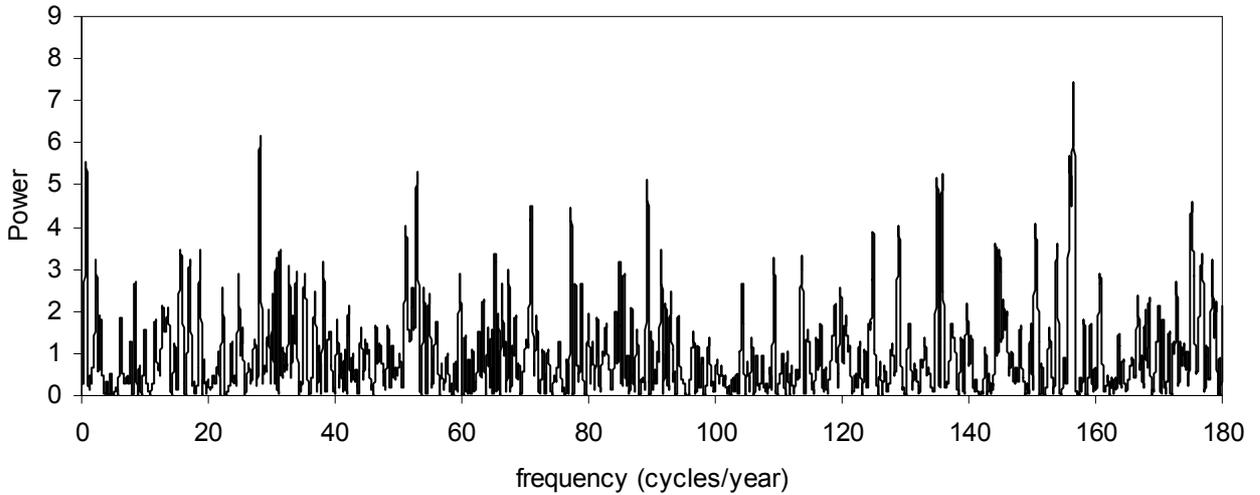

*Fig.4 – Salt phase Lomb-Scargle periodogram.*

Finally, in the combined spectrum in Fig. 5 the major peaks are 7.54 (150.66), 6.13 (36.84), 5.99 (70.76) and 5.87 (118.56).

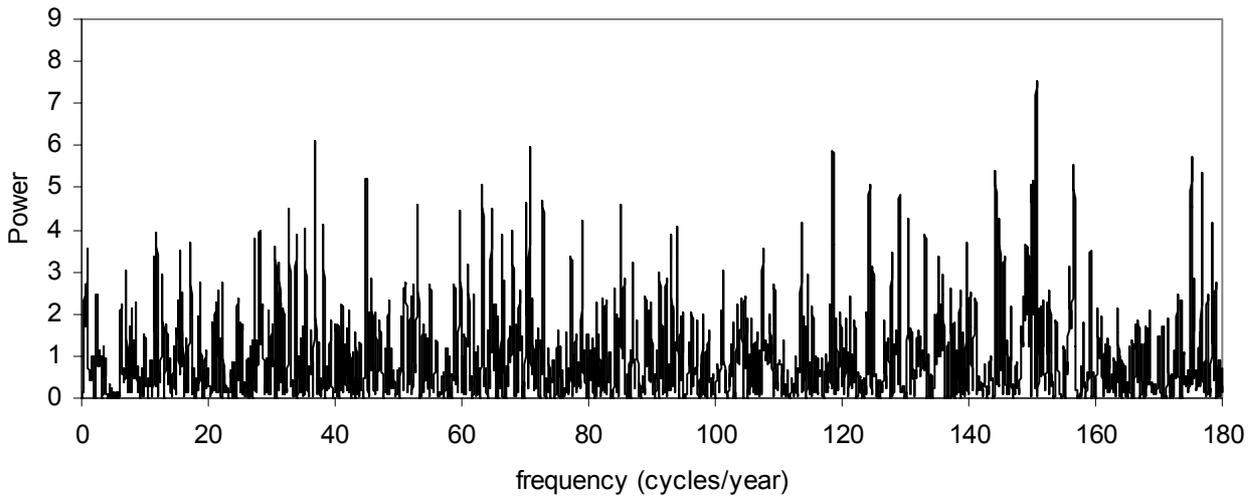

*Fig.5 – Combined dataset Lomb-Scargle periodogram.*

The discussion of the consistency of these spectra with the hypothesis of constant rate is postponed to the next paragraph IV, where the issue of the significance of the major peaks identified in each of them will be addressed following the Monte Carlo procedure described before.

**B. Likelihood Generalization of the Lomb-Scargle Method**

*1. Methodology and statistical properties*

In various works (for example [12], [8]) it has been explained how the standard Lomb-Scargle search can be generalized via a likelihood method that allows to take into account fully the probabilistic nature of the underlying process generating the series under study, i.e. in the present case the Poisson nature of the counting process.

For the problem being considered such an extension can be performed as follows; by denoting the variable average neutrino flux as

$$\mu(t) = \mu_{\cos t}[1 + a \cdot sen(\omega t + \varphi)] \quad (8)$$

we have that the expected mean rate for the $r_{th}$ bin, represented by the weighted mean time $t_r$, can be written as

$$\mu_r = \mu(t_r)\Delta t_r \quad (9)$$

where $\Delta t_r$ is the width of the bin itself; the likelihood for the detected sequence of values is consequently

$$L = \prod_{r=1}^{N} \frac{\mu_r^{n_r} e^{-\mu_r}}{n_r!} \quad (10)$$

$n_r$ being the number of counts in the $r_{th}$ bin. $\mu_{cost}$ is the average count rate evaluated separately for the D$_2$O and salt phases, and $a$ and $\varphi$ are the amplitude and phase maximization parameters.

Resorting to the Wilks' theorem (or log-likelihood ratio theorem) [13] the generalized likelihood ratio can be written as

$$GLR = \frac{\prod_{r=1}^{N} \frac{\mu_{\cos t}^{n_r} e^{-\mu_{\cos t}}}{n_r!}}{\max_{a,\varphi} \prod_{r=1}^{N} \frac{\mu_r^{n_r} e^{-\mu_r}}{n_r!}} \quad (11)$$

with the theorem stating that the quantity $-\ln(GLR)$

$$-\ln(GLR) = S(\omega) = \max_{a,\varphi} \sum_{r=1}^{N}(-\mu_r + n_r \cdot \ln \mu_r) - \sum_{r=1}^{N}(-\mu_{\cos t} + n_r \cdot \ln \mu_{\cos t}) \quad (12)$$

is exponentially distributed as $e^{-S}$ under the null hypothesis; $S=-\ln(GLR)$ is by definition the likelihood spectrum of the process. The likelihood spectrum thus shares the same property of the Lomb-Scargle periodogram that the spectral ordinate at each frequency in case of pure noise follows an exponential distribution. As a consequence, it can be presumed that the distribution of the highest peak follows eq. (5) and that more in general the distributions of the peaks ranked in term of height should follow the model (6).

Clearly, also in the framework of the likelihood method a Monte Carlo approach similar to that outlined for the Lomb-Scargle case can be adopted to assess the significance of the largest peaks. In the relevant studies reported in paragraph IV it will be checked how faithfully the formulas (5) and (6) are able to reproduce the actual Monte Carlo histograms.

*2. Experimental spectra*

The spectra computed according to the relation (12) for the D$_2$O phase, the salt phase and the combined dataset are shown in Fig. 6, 7 and 8.

In the spectrum in Fig. 6 the four highest peaks have, respectively, ordinates 7.53 (149.16), 5.58 (118.53), 4.8 (150.08) and 4.55 (101.19) (again in parenthesis there are the corresponding frequencies in cycles/year).

In the spectrum in Fig. 7 the four highest peaks have, respectively, ordinates 6.18 (156.53), 6.01 (52.91), 5.46 (28.16) and 5.04 (0.75).

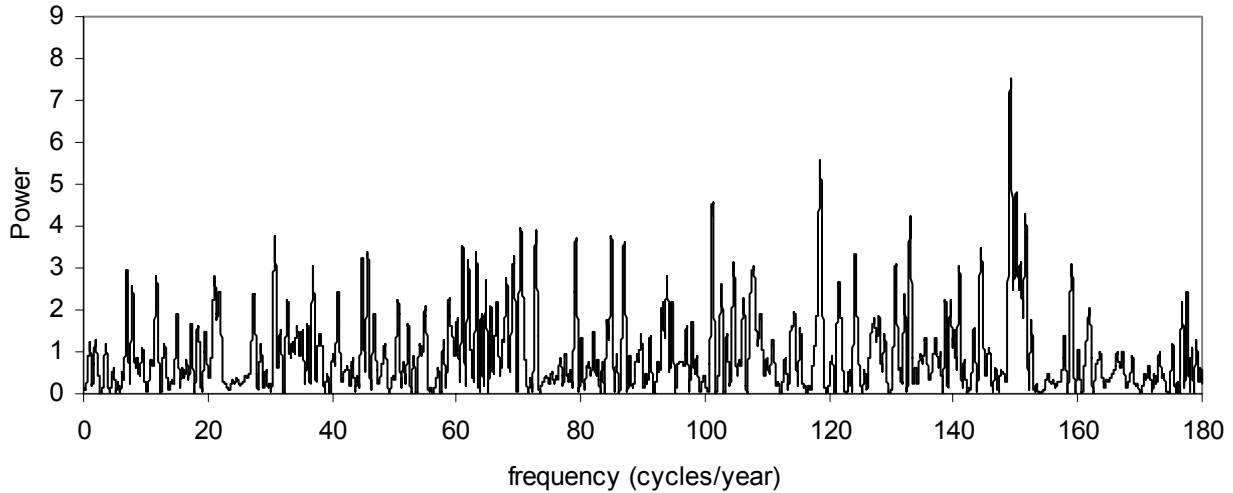

*Fig.6 – $D_2O$ phase likelihood spectrum.*

Finally, in the spectrum in Fig. 8 related to the combined dataset the four highest peaks have ordinates 6.84 (150.67), 5.85 (36.84), 5.41 (144.24) and 5.31 (53.03).

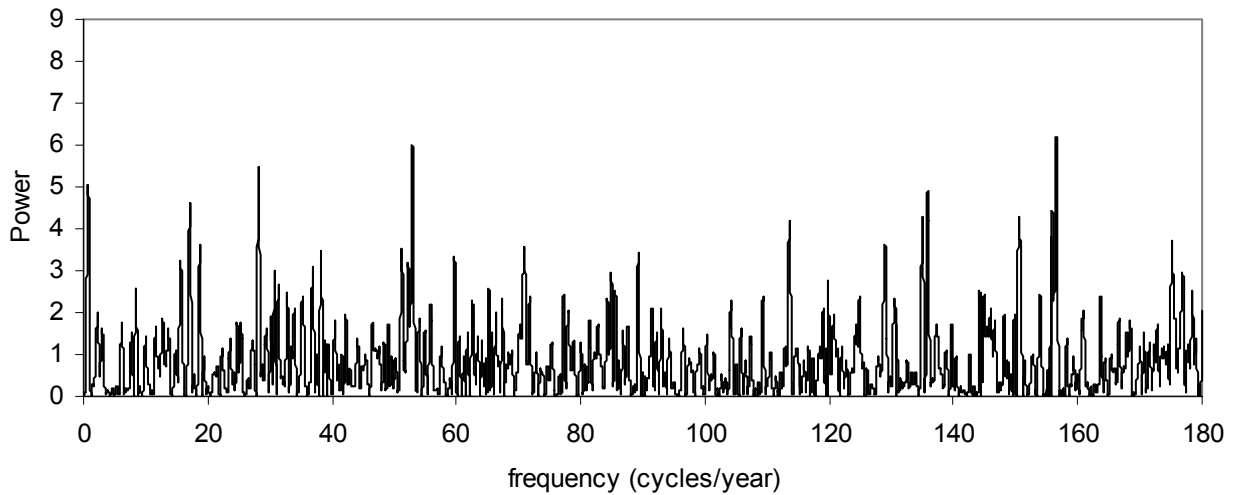

*Fig.7 – Salt phase likelihood spectrum.*

We move now to the significance assessment via the extensive Monte Carlo studies reported in the next paragraph.

## IV. NULL HYPOTHESIS DISTRIBUTIONS AND SIGNIFICANCE ASSESSMENT

All the reference null hypothesis Monte Carlo distributions reported in this paragraph have been inferred from 10000 simulation cycles; in each cycle a synthetic series is generated and the relevant spectrum is computed. Specifically, each ordinate value of the synthetic series is generated by drawing a random number from a Poisson distribution with mean value equal to the product of the experimental average rate times the live time of the corresponding bin, and then by dividing the drawn number by the live time itself. In this way the simulated series automatically retain timing and statistical properties of the experimental series.

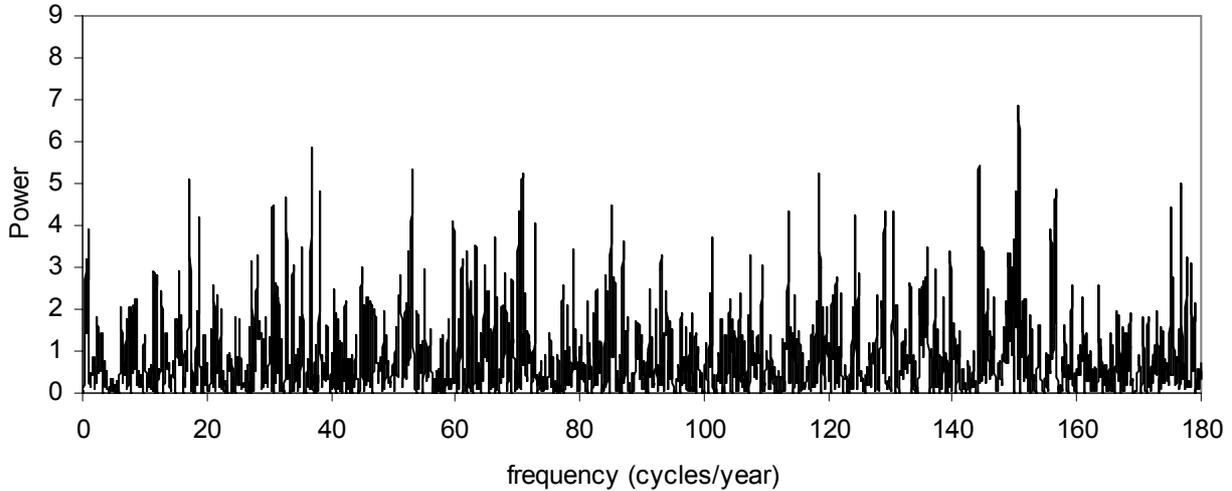

*Fig.8 – Combined dataset likelihood spectrum.*

In the following, after describing the features of the simulated distributions, the significances of the major spectral peaks are listed both for the Lomb-Scargle and likelihood method. As consistency check, it must be added that we verified throughout the simulation process how precisely the individual frequencies of the simulated spectra follow the predicted exponential distribution, finding a perfect agreement between the simulation results and the theoretical expectations for both methodologies.

**A. Lomb-Scargle Distributions**

The Lomb Scargle null hypothesis distributions for the $D_2O$ phase, salt phase and the combined dataset are reported in Fig. 9, 10 and 11, respectively. In each figure there are the distributions of the four highest peaks, as well as the plots of the model functions (6).

It is worth to point out that the model functions are normalized to unit area, and thus for comparison purpose also the Monte Carlo curves have been normalized in the same way.

It can be noted that the Monte Carlo distributions of the peaks (the analysis is limited to the four highest peaks) follow remarkably well the model functions with a number of effective independent frequencies equal to 752 for the $D_2O$ dataset, 936 for the salt dataset and 1876 for the combined dataset. Such numbers have been obtained through the fit to the model (5) of the distribution pertaining to the highest spectral peak.

A slight disagreement, however, can be noted for the curves related to the third and fourth peaks. Even though it can be shown [8] that the perfect agreement of the simulated curves to the general model (6) is expected only in the case of evenly sampled time series, when the Lomb-Scargle periodogram reduces to the so called Schuster periodogram [7], the good agreement of the model to the Monte Carlo distributions in the Lomb-Scargle case is not a surprise, since it was already demonstrated via extensive simulations in [3], at least for the highest peak.

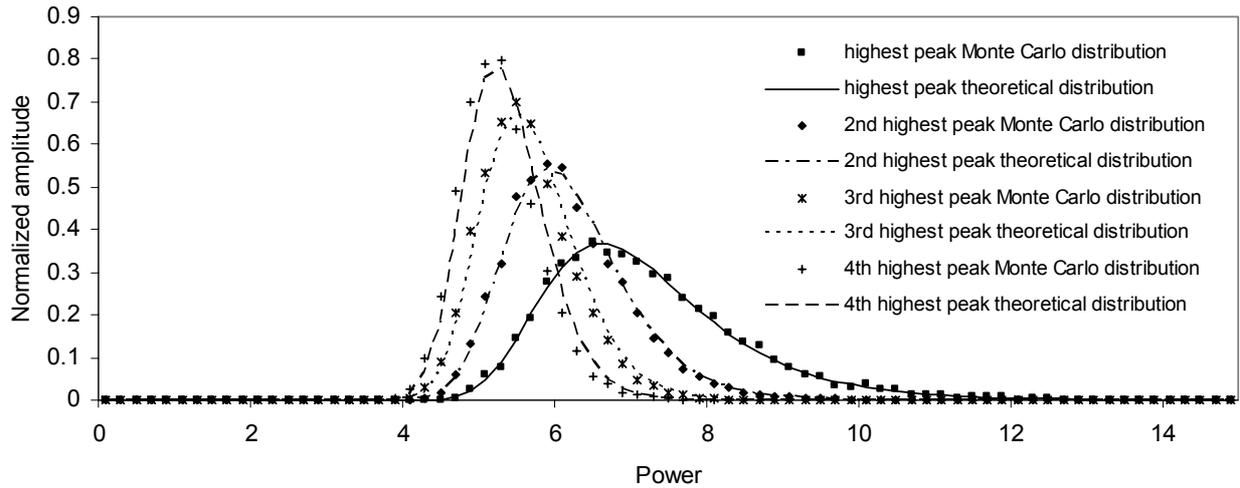

*Fig. 9 – D$_2$O phase null hypothesis distributions for the Lomb-Scargle method. The overlapped curves are the model functions (6) plotted with M=752.*

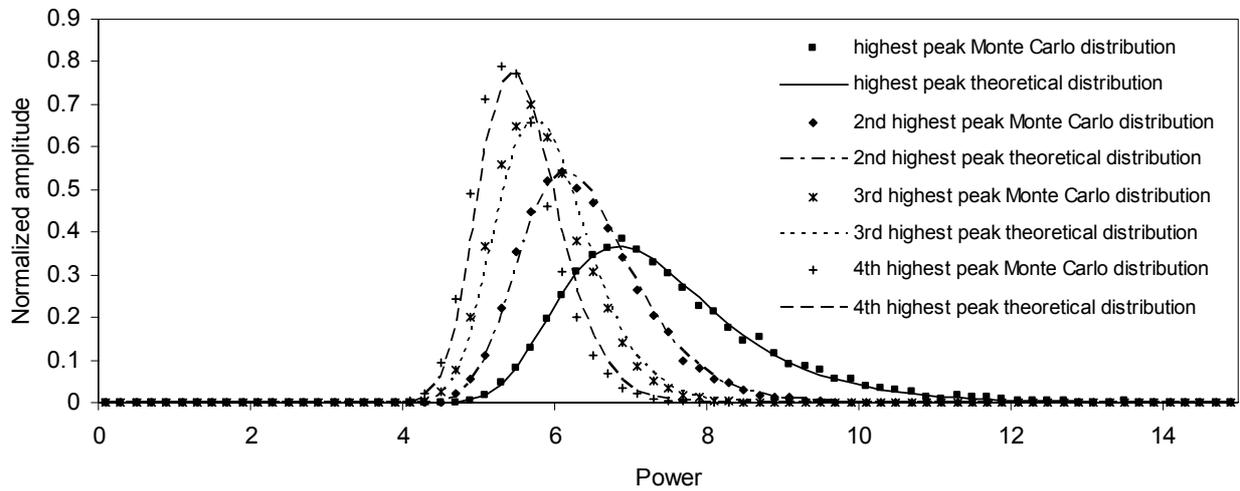

*Fig. 10 – Salt phase null hypothesis distributions for the Lomb-Scargle method. The overlapped curves are the model functions (6) plotted with M=936.*

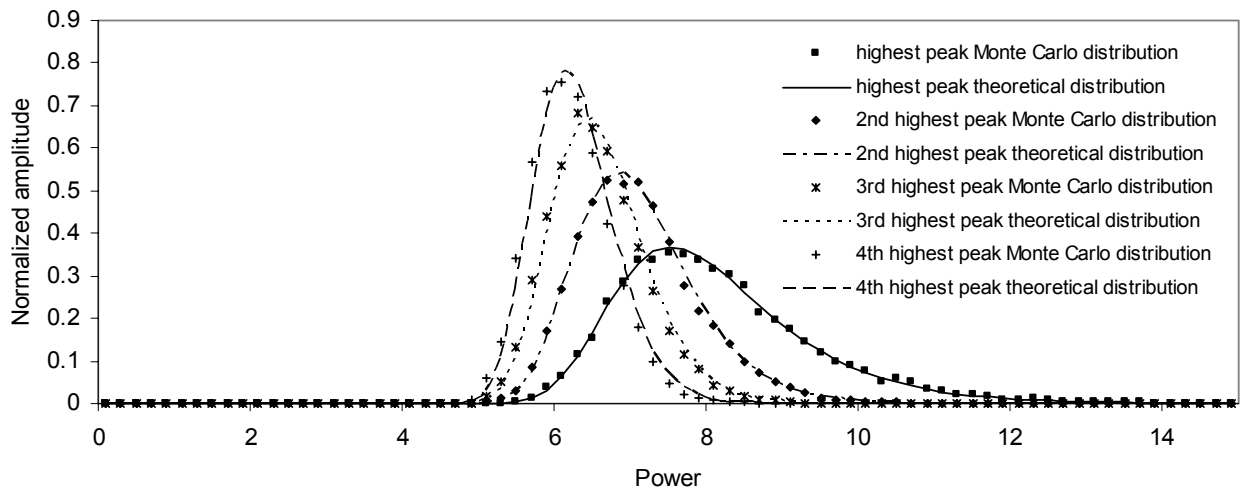

Fig. 11 – *Combined dataset null hypothesis distributions for the Lomb-Scargle method. The overlapped curves are the model functions (6) plotted with M=1876.*

| Ordinate | Frequency | Significance |
|---|---|---|
| 8.11 | 149.13 | 20.7% |
| 5.59 | 118.60 | 75.5% |
| 5.55 | 44.86 | 51.7% |
| 5.54 | 150.03 | 27.7% |

Table 1 -. D$_2$O Lomb-Scargle significance.

| Ordinate | Frequency | Significance |
|---|---|---|
| 7.43 | 156.52 | 42.9% |
| 6.16 | 28.15 | 56.4% |
| 5.55 | 0.75 | 65.9% |
| 5.32 | 52.93 | 58.9% |

Table 2 -. Salt Lomb-Scargle significance.

| Ordinate | Frequency | Significance |
|---|---|---|
| 7.54 | 150.66 | 63.4% |
| 6.13 | 36.84 | 90.7% |
| 5.99 | 70.76 | 82.1% |
| 5.87 | 118.56 | 72.2% |

Table 3 -. Combined data set Lomb-Scargle significance.

As illustrated before, from the simulated distributions we can get the significance of the four highest peaks of each spectrum; the results are summarized in the tables 1, 2 and 3, respectively. In no cases there are "interesting" low or very low significance levels, and hence the values in the tables taken all together show unambiguously that the three spectra are very well consistent with the hypothesis that the SNO data time series are due to a constant rate process.

**B. Distributions for the Likelihood Generalization of the Lomb-Scargle Method**

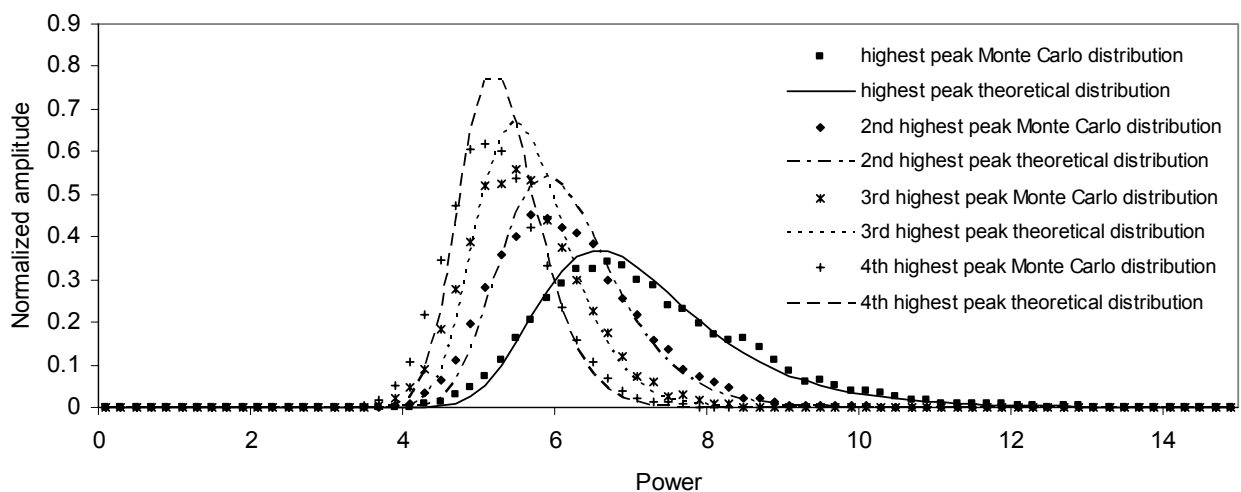

*Fig. 12 – D$_2$O phase null hypothesis distributions for the likelihood method. The overlapped curves are the model functions (6) plotted with M=724.*

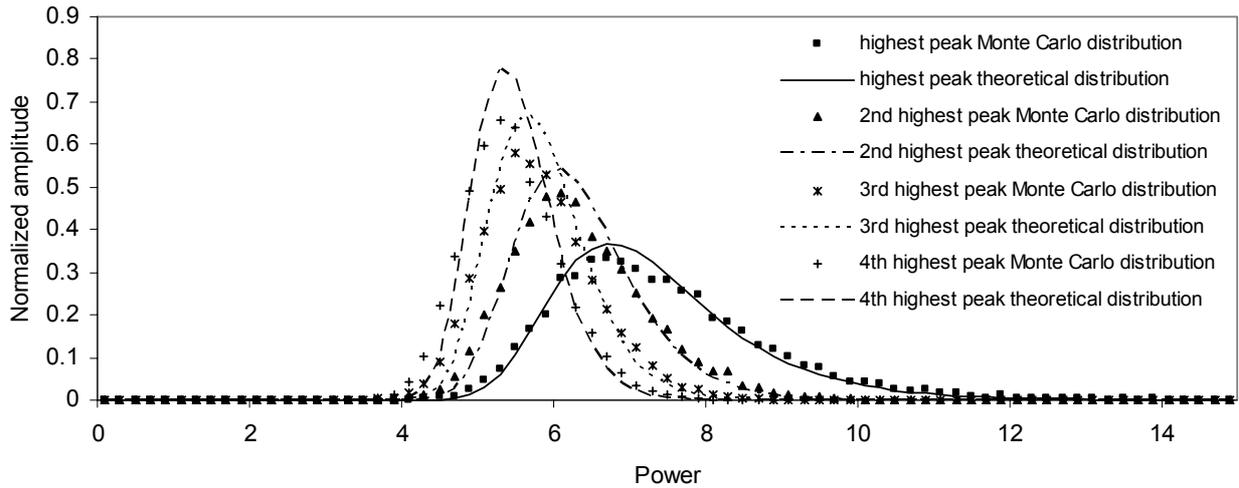

*Fig. 13 – Salt phase null hypothesis distributions for the likelihood method. The overlapped curves are the model functions (6) plotted with M=856.*

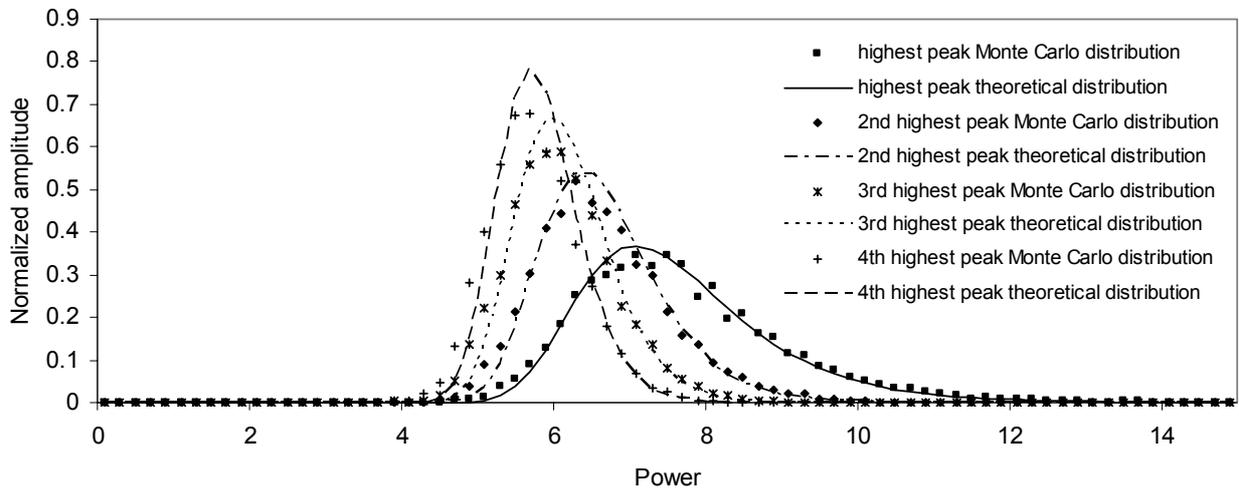

*Fig. 14 – Combined dataset null hypothesis distributions for the likelihood method. The overlapped curves are the model functions (6) plotted with M=1196.*

The null hypothesis distributions for the $D_2O$ phase, salt phase and the combined dataset stemming from the application of the likelihood generalization of the Lomb-Scargle method are reported in Fig. 12, 13 and 14. Also in these figures there are the distributions of the four highest peaks overlapped to the respective model functions (6).

It can be noted that now the peak distributions follow only approximately the model functions, far from the good agreement observed in the Lomb-Scargle case. Specifically, the value $M$ obtained through the approximate fit of the highest peak distribution to the equation (5) equals 724, 856 and 1196 in the $D_2O$, salt and combined datasets, respectively. The significance of the four major peaks in each spectrum, inferred from the corresponding null hypothesis distributions, are summarized in the tables 4, 5 and 6. By examining the values in the tables it can be immediately concluded that also this alternative methodology demonstrates that all the three spectra are fully consistent with the hypothesis of a constant rate time series.

| Ordinate | Frequency | Significance |
|---|---|---|
| 7.54 | 149.16 | 34.5% |
| 5.58 | 118.53 | 71.5% |
| 4.80 | 150.08 | 87.5% |
| 4.55 | 101.19 | 86.9% |

Table 4 -. D$_2$O Likelihood significance.

| Ordinate | Frequency | Significance |
|---|---|---|
| 6.18 | 156.53 | 81.7% |
| 6.02 | 52.91 | 60.6% |
| 5.46 | 28.16 | 66.1% |
| 5.04 | 0.75 | 73.4% |

Table 5 -. Salt Likelihood significance.

| Ordinate | Frequency | Significance |
|---|---|---|
| 6.84 | 150.67 | 71.7% |
| 5.85 | 36.84 | 82.2% |
| 5.41 | 144.24 | 84.8% |
| 5.31 | 53.03 | 76.3% |

Table 6 -. Combined Likelihood significance

## V. SENSITIVITY OF THE METHODS TO TRUE PERIODICITIES

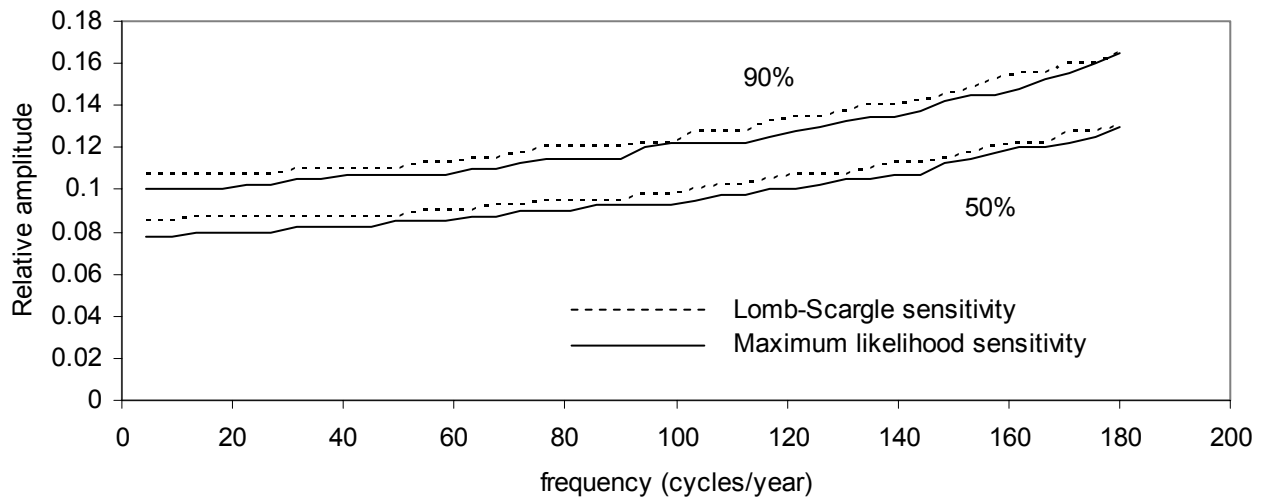

*Figure 15 – Sensitivity contours for the Lomb-Scargle method and for its likelihood extension. The contours are the loci on the amplitude-frequency plane at which there is a 50% (90%) probability to discover a true sinusoidal signal at the 99% of confidence level. The likelihood method performs marginally better than the Lomb-Scargle approach.*

The sensitivity of the two methods under consideration to a true periodicity can be evaluated by defining a desired confidence level for a discovery claim, for example 99%, which is equivalent to fix a detection threshold on the null hypothesis distribution producing a false detection only in 1% of the cases; then via Monte Carlo it is determined for each frequency the modulation amplitude that, above the selected threshold, ensures a detection at a prescribed level of probability, for example 50% and 90%. The noise distribution to consider to set the detection threshold is clearly that relevant to the highest spectral peak.

The 50% and 90% sensitivity contours for the confidence level of 99% are shown in Fig. 15 for both the Lomb-Scargle and likelihood methods. It can be noted that the sensitivities of the two approaches are pretty similar, the likelihood method being only marginally better then the Lomb-Scargle one.

Such curves can be compared with those reported in [1]. In particular, apart from the different scale on the $x$ axis, the contours for the Lomb-Scargle method computed here appear to be close to those for the weighted Lomb-Scargle method reported in the figure 4 of [1]. Furthermore, it is also confirmed that at low frequency, where the likelihood approach used here should coincide with the unbinned likelihood of [1], an 8% amplitude modulation would be discovered 50% of the times at the confidence level of 99%.

The comparison of the two methods described here and of the two adopted in [1] demonstrates that at low frequency the intrinsic statistical power of the methods is essentially the same. The superiority of the unbinned maximum likelihood used by the SNO Collaboration is in that it deals automatically better with the frequency, thus assuring a constant behaviour of the detection sensitivity as function of the frequency itself.

## VI. SPECIFIC FREQUENCIES

### A. Annual Modulation

It may be interesting to check how the annual modulation due to the Earth's orbit eccentricity is recovered from the data through the previous methodology. We consider for this case the results of the likelihood method for the combined dataset: the amplitude is evaluated to be 0.0337±0.029 and the phase 0±1.103. In Fig. 16 the modulation described by the fitted amplitude and phase values is compared to that geometrically expected; the amplitude is picked up well by the fit, even though with a large error. For what concern the phase, there is a shift between the best fit and the expected curve of about 40 days.

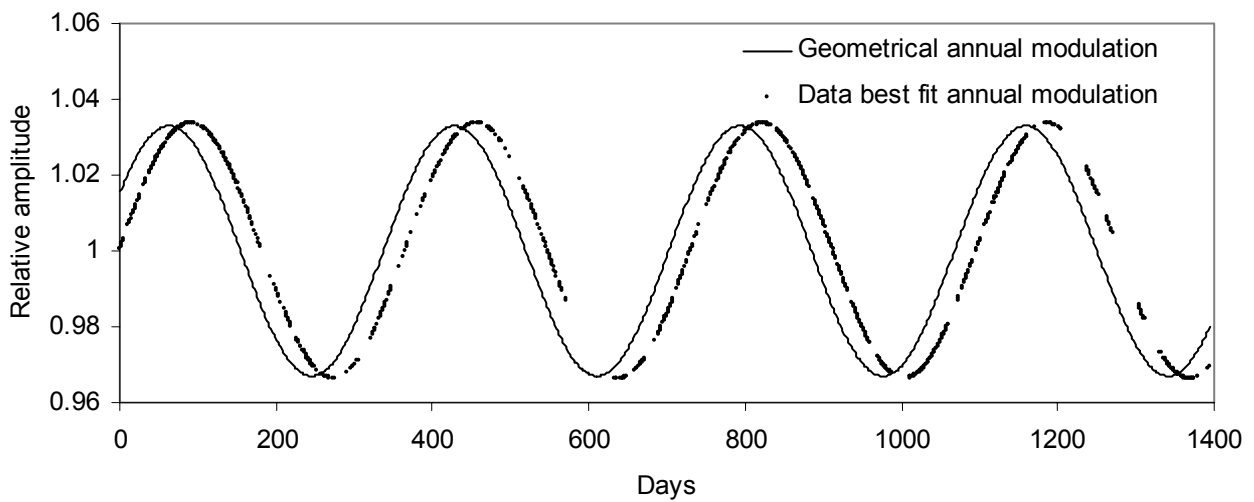

*Figure 16 – Annual modulation: geometrical expectation vs. the best fit inferred from the combined dataset.*

The spectral amplitude at this frequency is 2.26 and thus clearly well below the discovery capability of both methods. This fact is very clearly shown in Fig 17, where the expected spectral distribution of the annual modulation line of 3.3% relative amplitude is reported together with the global and single frequency noise distributions. The global noise distribution, in particular, masks completely the expected signal distribution, thus virtually preventing any detection possibility in a "blind" search over the entire investigated frequency range.

The situation is better if one considers only the frequency of interest; but even in this case, by determining the detection threshold that would ensure on the exponential noise distribution a false alarm probability of 1%, the resulting detection probability would be only about 14%. So, in other words, without knowing that such a modulation must be there, it could not be recognized. This is clearly not surprising, since the modulation amplitude is pretty low if compared with the Poisson scattering of the data.

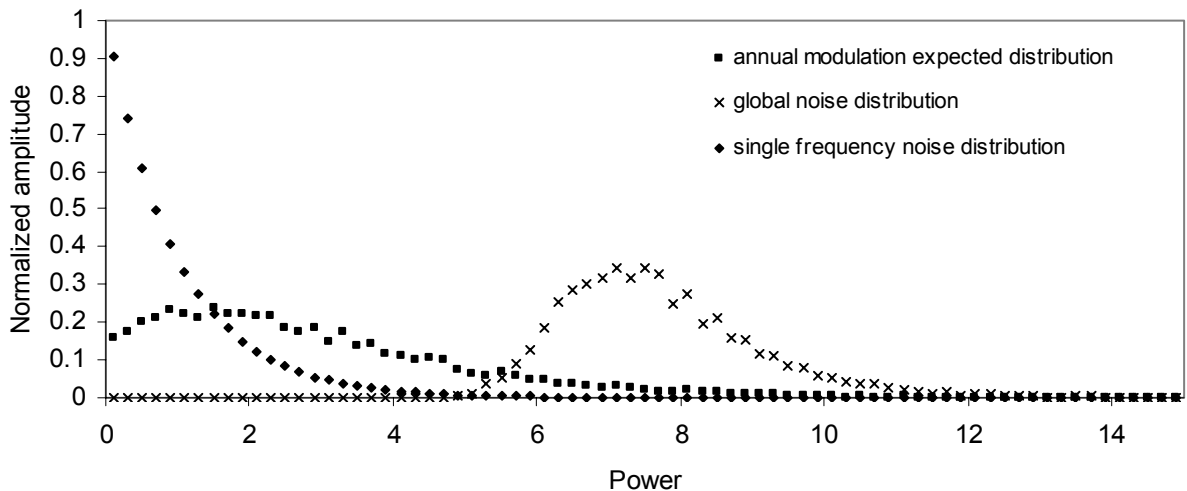

*Figure 17 – Annual modulation: expected spectral amplitude distribution.*

### B. 9.42 cycles/year Frequency

Recent analyses of the Super-Kamiokande I data [8][14] pointed toward a spectral feature at 9.42-9.43 cycles/year, with controversial significance, that, if interpreted as a true signal, would correspond to a modulation of relative amplitude between 6 and 7%. It is thus interesting to analyze in this respect the SNO data.

In Fig. 18 it is shown the expected amplitude distribution of the corresponding spectral ordinate assuming a true amplitude of 6.5 %, compared with the single frequency and global noise distributions. The broad signal distribution is largely masked by the noise distribution. Thus, also in such a case the "blind" search would hardly lead to discover the signal if present, but, assuming the a-priori knowledge of the frequency to be searched, the detection efficiency would be quite high; indeed fixing as above the 99% confidence level threshold on the exponential single frequency noise distribution, the corresponding detection probability would result equal to about 82.6%.

In the combined likelihood spectrum in Fig.8, in the region between 9.32-9.52 cycles/year the maximum ordinate is 0.407 at the frequency of 9.41 (the corresponding amplitude best fit value is $0.0128^{+0.0228}_{-0.0128}$). This spectral value appears immediately inconsistent with the expected signal distribution in Fig. 18, since it lies on its extreme left tail region. Quantitatively, the Monte Carlo result gives a probability to get a value as low or lower of 0.407, given a signal actually present, less than 0.1 percent.

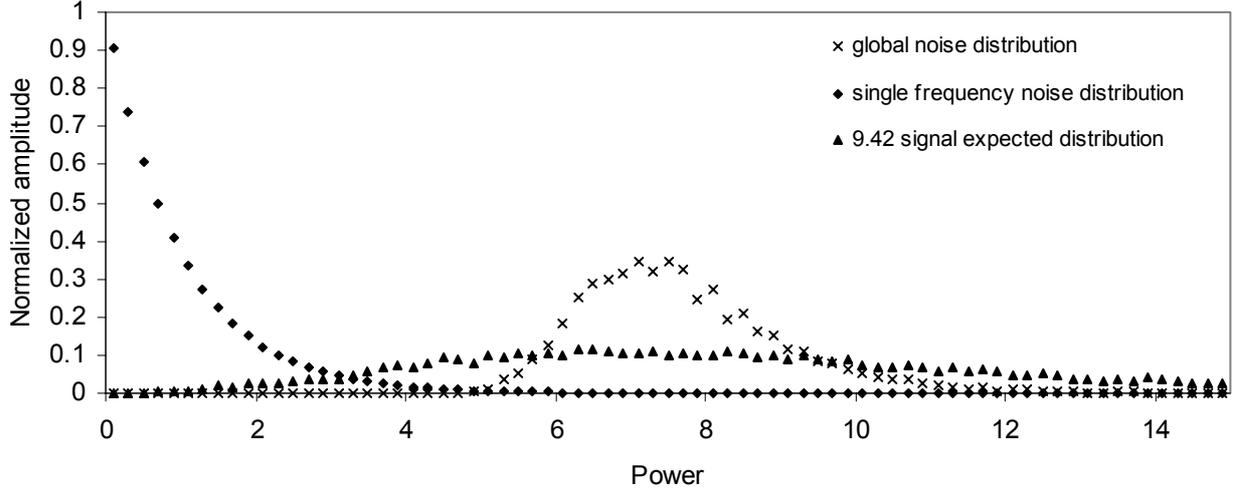

*Fig. 18 – 9.42 cycles/year signal: expected spectral amplitude distribution.*

In summary, the SNO data are, thus, strongly in disagreement with the presence of a 9.42-9.43 cycles/year modulation, in the 6-7% amplitude range, in the measured $^8$B solar neutrino flux.

## VII. HOW TO APPROXIMATE THE UNBINNED LIKELIHOOD APPROACH

The digression in this paragraph is intended to illustrate a methodology which exploits the released 1 day truncated data in a way suited to approximate the unbinned likelihood approach adopted by the SNO Collaboration. Essentially what it is proposed here is to write down a sort of "1 day binned" likelihood which takes into account the data taking period of each 1 day bin as it is, without thus resorting to the concept of weighted mean time adopted in the methods extensively developed in the previous paragraphs.

To this purpose, the actual number of counts measured in each bin is supposed drawn from a Poisson distribution with parameter $\mu_r$ given by

$$\mu_r = \sum_{r_k=1}^{R} \int_{t_{sr_k}}^{t_{er_k}} \mu_{\cos t}[1 + a \cdot sen(\omega t + \varphi)] \, dt \tag{13}$$

where R is the number of data taking intervals in the $r_{th}$ 1 day bin and $t_{srk}$ and $t_{erk}$ are the corresponding start and end times.

Following the notation used for the eq. (12), the "1 day binned" likelihood spectrum hence becomes

$$S(\omega) = \max_{a,\varphi} \sum_{r=1}^{N}(-\mu_r + n_r \cdot \ln \mu_r) - \sum_{r=1}^{N}(-\mu_{\cos t} + n_r \cdot \ln \mu_{\cos t}) \tag{14}$$

that, because of the Wilks' theorem, is exponentially distributed as $e^{-S}$ under the null hypothesis.

Due to the intensive computing requirements of such a methodology, in the following only the case of the combined dataset will be considered, whose spectrum according eq. (14) is shown in Fig. 19.

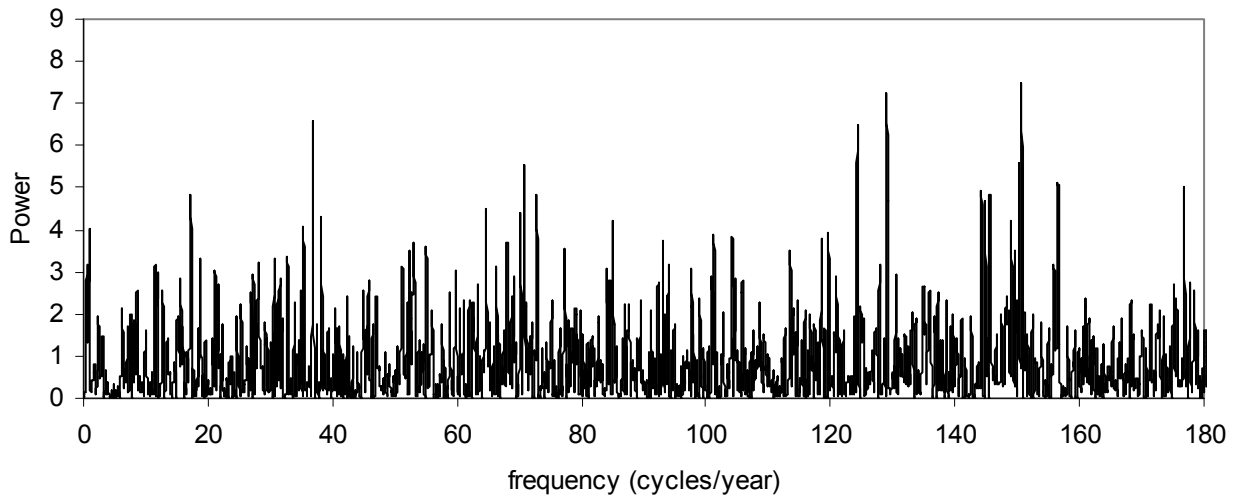

*Fig.19 – Combined dataset "1 day binned" likelihood spectrum.*

In the spectrum the four highest peaks have, respectively, ordinates 7.46 (150.69), 7.23 (129.05), 6.60 (36.84) and 6.51 (124.31) (as usual in parenthesis there are the corresponding frequencies in cycles/year).

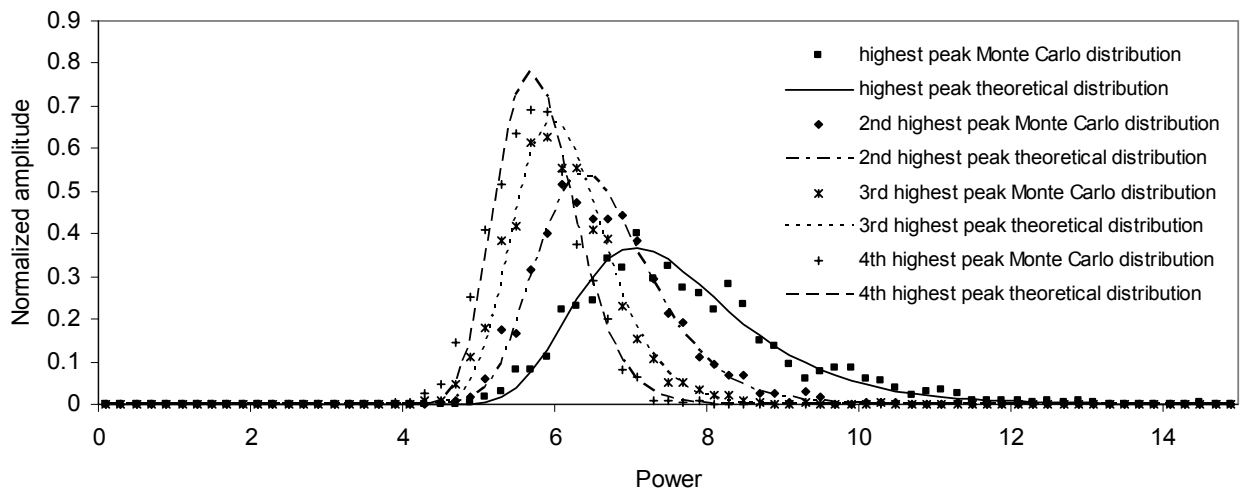

*Fig. 20 – Combined dataset null hypothesis distributions for the "1 day binned" likelihood method. The overlapped curves are the model functions (6) plotted with M=1182.*

The corresponding null hypothesis distributions, evaluated trough a set limited to 1000 simulations because of the computing constraints mentioned above, are reported in Fig. 20. Also in this case the peak distributions follow only approximately the model functions, far from the good agreement observed in the Lomb-Scargle case, with an *M* value, obtained through the approximate fit of the highest peak distribution to the equation (5), equal to 1182. From such distributions it can be inferred that the significance of the highest peak is 49.7%, of the second highest is 22.1%, of the third highest 22.0%, and finally of the fourth highest 9.7%. Therefore, also such methodology produces a spectrum perfectly consistent with the constant rate hypothesis, further confirming the results obtained with the previous analyses.

Finally, in Fig. 21 the sensitivity plot of Fig. 15 is updated with the addition of the sensitivity contours stemming from this more refined methodology; as expected, at low frequency the sensitivity of the 50% and 90% contours reproduces that relevant to the approximate likelihood method, while at high frequency the new contours show an enhanced discovery capability, due to the more precise account of the actual run times within each nominal 1 day bin.

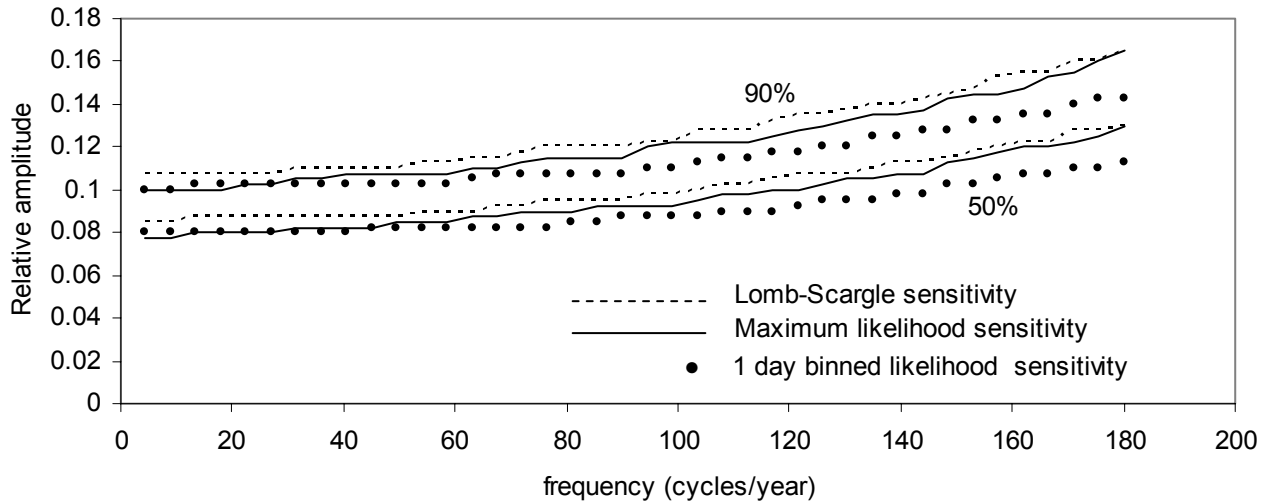

*Figure 21 – Sensitivity contours updated with the inclusion of the curves relevant to the more refined "1 day binned" likelihood methodology. Whilst there is no increase of sensitivity at low frequency, the new method outperforms the other two as the frequency increases.*

## VIII. JOINT ANALYSIS OF THE SNO AND SUPER-KAMIOKANDE I DATA OVER THE COMMON DATA TAKING PERIOD

The SNO data are characterized by an overlap period with the data of the water Cerenkov experiment Super-Kamiokande I; such a period ranges from November 2, 1999 to May 27, 2001, hence encompassing the whole $D_2O$ phase, while the salt phase is not involved at all. Since, as already mentioned in paragraph 6, the solar neutrino data from Super-Kamiokande have been carefully searched for possible periodicities, both by the Collaboration [5] and by other research groups [8][14], with controversial results, it can be worthwhile to attempt a joint periodicity search for possible common underlying neutrino flux modulations over the overlap data taking period, exploiting concurrently both the datasets.

Since the spectral line width depends essentially upon the reciprocal of the spanned time interval [15], the spectra of the two series, while computed only over the common data taking range, feature very similar resolutions, hence making it meaningful to sum each other. As a consequence, a simple and effective way to carry out the joint periodicity search could be to compute separately the spectra of the two series in the same time span, sum them, and then examine statistically the main lines of the spectrum resulting from the sum.

Among the different methods of analysis for the SNO time series illustrated in the first part of the paper, which proved to produce fairly comparable results, we choose here for the joint analysis the methodology referred as likelihood extension of the standard Lomb-Scargle approach. In [8], also for the Super-Kamiokande I dataset different approaches were illustrated to compute the spectrum: among them we adopt for the present study the so called asymmetric likelihood method, which ensures the more correct handling of the data errors.

The summed spectrum is reported in Fig. 22. It is plotted in the limited range up to 50 cycles/year, because the coarse binning (5-day bins) of the Super-Kamiokande data implies a lower overall upper limit for the frequency analysis with respect to the SNO data taken alone; specifically, the limit of 50 cycles/year is chosen for consistency with the previous SK analyses.

The four highest peaks have, respectively, ordinates 6.41 (45.68), 6.20 (21.16), 5.74 (44.76) and 5.64 (48.71) (as usual in parenthesis there are the corresponding frequencies in cycles/year).

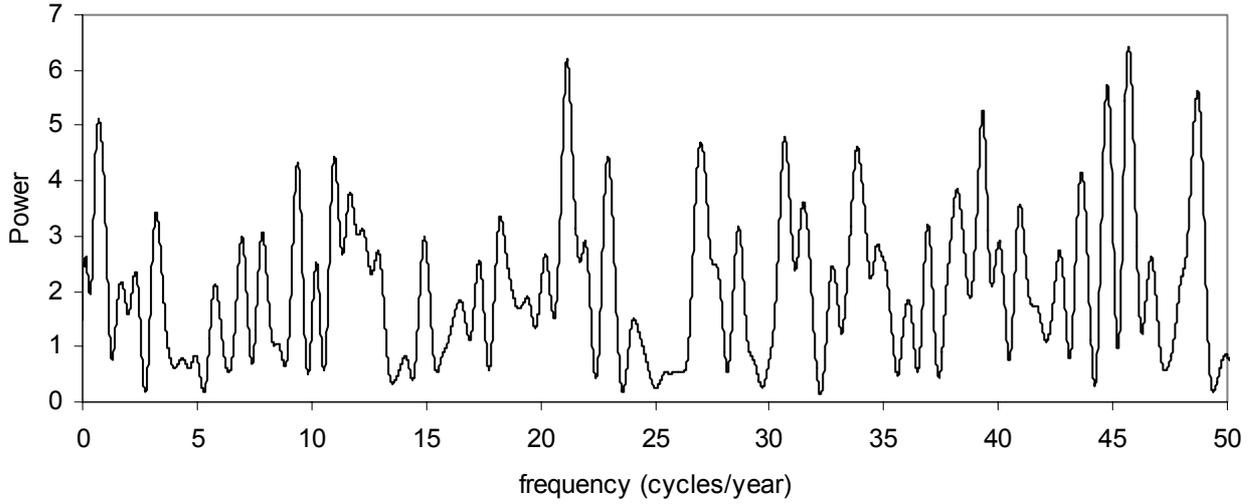

*Figure 22 – Spectrum resulting from the sum of the SNO and SK spectra computed in the overlap period from November 2, 1999 to May 27, 2001.*

We pass now to the evaluation of the null hypothesis distributions from which we can obtain the significances of the major lines in the summed spectrum. They are computed with a Monte Carlo approach which follows the guideline already illustrated in the first part of the paper: for each simulation cycle two random datasets are generated, one following the time and statistical features of the SNO $D_2O$ data and the other according to the characteristics of the Super-Kamiokande I data in the same time range. The datasets are generated under the null hypothesis, i.e. with no modulation embedded. The spectra of the two simulated series are added together, originating therefore 10000 simulated summed spectra, from which it is possible to infer the distributions of the highest spectral peaks.

Before discussing the significances as derived from the Monte Carlo, it can be instructive to remark how the statistical features of the summed spectrum are modified with respect to the original spectra. We know that, under the null hypothesis, the amplitude of each frequency of the original spectra is distributed according to the very simple exponential function $e^{-z}$; as a consequence, each frequency of the summed spectrum is distributed according to the convolution of two identical exponential functions, which can be very easily computed to be $ze^{-z}$. Adopting also here the concept of independently scanned frequencies, the distribution of the peaks of the summed spectra, ordered in terms of height, should hence follow the same relation (6), where *F(h)* is given by

$$F(h) = \int_0^h p(\lambda)d\lambda \qquad (15)$$

Where now *p(h)* is $he^{-h}$.

In order to show how such modified, expected distribution features are reflected in the Monte Carlo output, we plot in Fig. 23 the single frequency distribution and the highest peak distribution for the summed spectrum, as stemming from 10000 simulation cycles. In particular, the agreement of the former distribution with the theoretical $ze^{-z}$ function is extremely good, thus giving confidence about the reliability both of the model and of the Monte Carlo. On the other hand, the Monte Carlo distribution of the highest peak follows only approximately the theoretical model (6), plotted for a value of M equal to 175, therefore showing that also in the present case the concept of independent scanned frequencies has to be regarded only as an approximation.

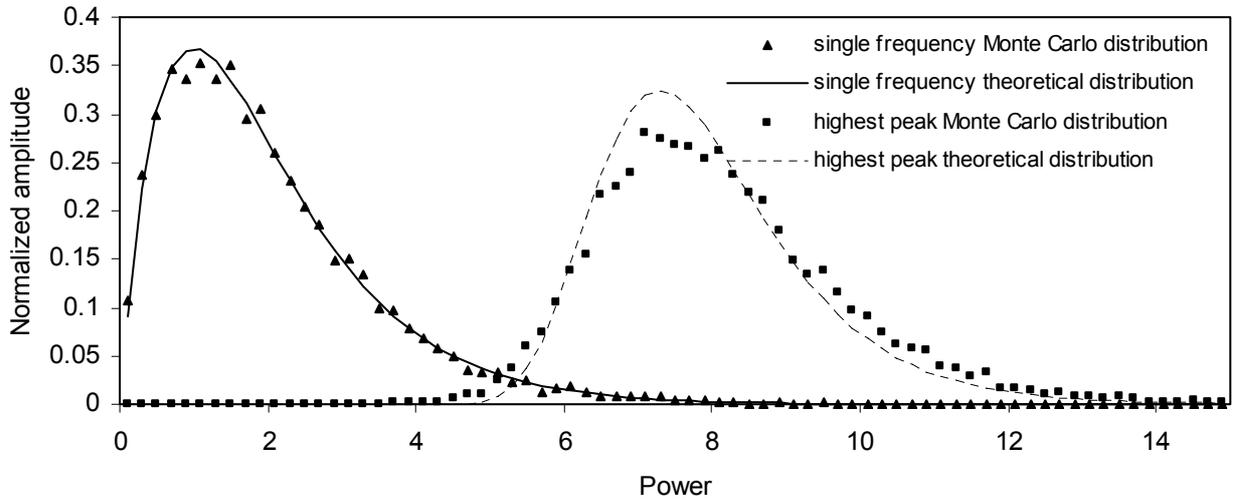

*Fig 23 – Comparison of the theoretical single frequency and highest peak spectral distributions with the same quantities derived from the Monte Carlo, for the SK and SNO summed spectrum (overlap period). As expected, for the former distribution the agreement between the theory and the Monte Carlo is very good.*

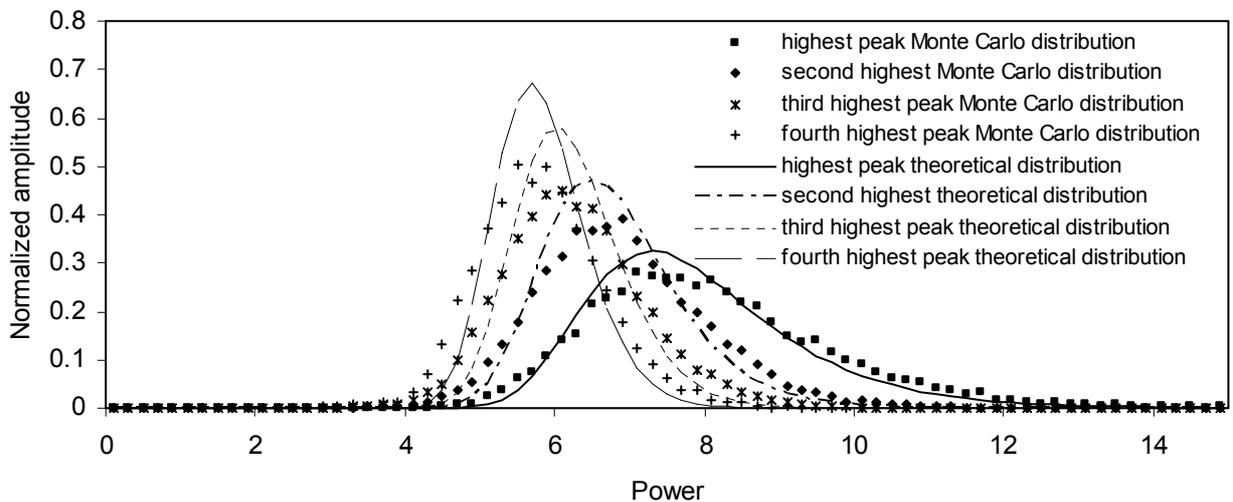

*Fig 24 – Summed spectrum null hypothesis distributions for the SK and SNO data in the common data taking period. The overlapped curves are the model functions (6) plotted with M= 175.*

The null hypothesis distributions for the first four highest peaks, together with the corresponding model functions plotted for M=175, are shown in Fig. 24; from them we get the significance values, which for the highest peak is 87.2 %, for the second highest is 72.3%, for the third highest 70.1%, and for the four highest 56.1 %.
In conclusion, the summed spectrum is fully consistent with the hypothesis of constant rate of the solar neutrino flux over the common measurement period of the two experiments.

To conclude this analysis, it is useful to confirm numerically that the potential discovery capability of a real modulation of the neutrino flux, if present in the commonly spanned interval, is higher for the summed spectrum with respect to the two individual spectra taken alone.

For this purpose we construct the 90% (Fig. 25) and 50% (Fig. 26) sensitivity contours for the three cases, i.e. the summed spectrum and the two individual spectra, computed for a false alarm probability value of 1% (i.e. 99% confidence level). Clearly the discovery capability of the former outperforms over the entire range that of the other two, even if, as it should be expected, the beneficial effect of the sum vanishes at the end of the spanned frequency interval because of the vanishing sensitivity of the SK data to frequencies too high with respect to their 5-day binning.

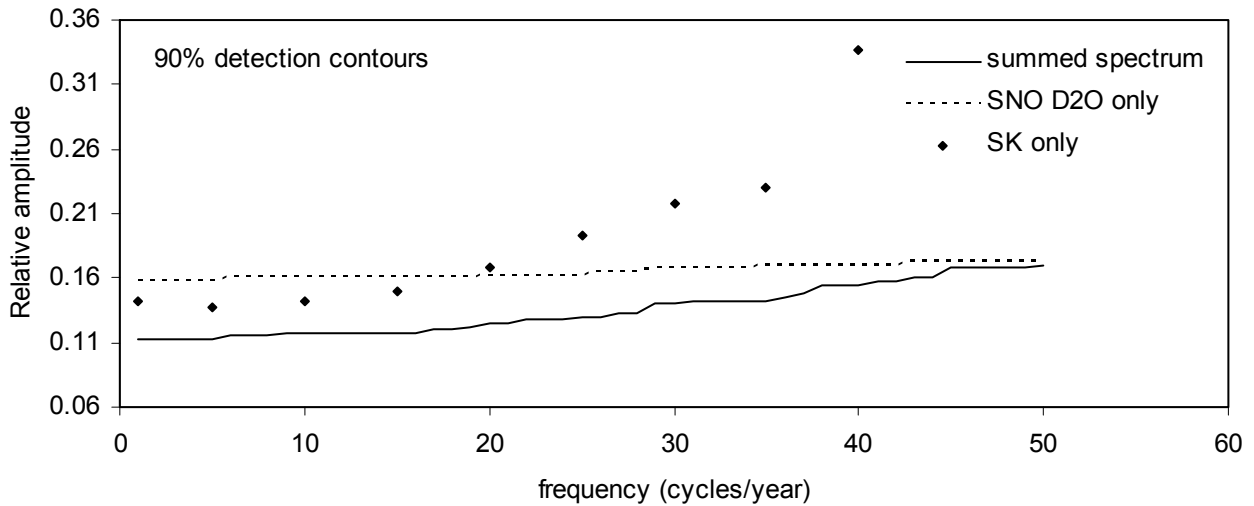

*Figure 25 – 90% sensitivity contour at the 99% confidence level for the summed spectrum, compared with the same contours inferred from the SK data alone and the SNO data alone (all the data refer to the common data taking interval). The summed spectrum performances outperform those of the two spectra considered separately.*

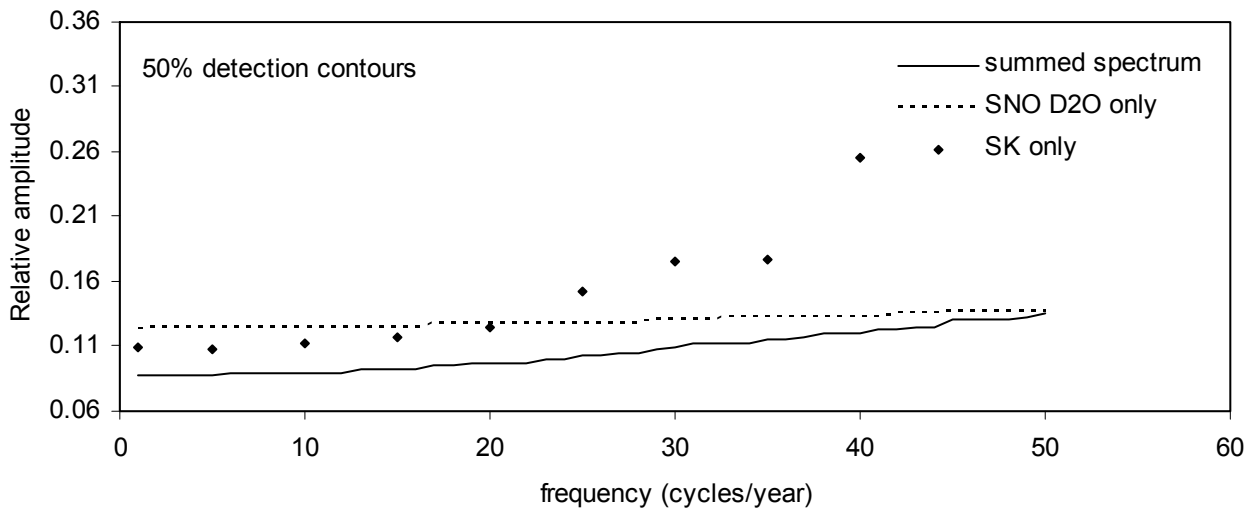

*Figure 26 – 50% sensitivity contour at the 99% confidence level for the summed spectrum, compared with the same contours inferred from the SK data alone and the SNO data alone (all the data refer to the common data taking interval). The previous result that the summed spectrum ensures better performances than the two original spectra taken separately is confirmed.*

## IX. CUMULATIVE ANALYSIS OF THE SNO AND SUPER-KAMIOKANDE I DATA INCLUDING ALL MEASUREMENT PERIODS

Taking together all the SNO and Super-Kamiokande I data, instead of only those pertaining to the commonly spanned time interval, a global analysis over the whole data taking period can be carried out by constructing their cumulative likelihood spectrum. In principle, to write down the formulation of such a spectrum it is enough to multiply the generalized likelihood ratio (11), valid for the SNO data, with the analogous quantity applicable to the Super-Kamiokande I data, i.e. eq. (48) of [8]: minus the logarithm of the global generalized likelihood ratio obtained in this way is, by definition, the cumulative spectrum of the two series.

In doing this, for the SNO data we consider the generalized likelihood ratio written according to the methodology which extends the Lomb-Scargle approach, while for the SK data we resort to the treatment with the explicit account of the asymmetric errors. Hence, the resulting formulation of the cumulative spectrum is

$$S(\omega) = \left( \frac{1}{2} \sum_{k=1}^{N_{SK}} \frac{x_k^2}{\sigma_{kp1}^2} - \sum_{r=1}^{N_{SNO}} (-\mu_{\cos t} + n_r \cdot \ln \mu_{\cos t}) \right) +$$

$$+ \max(a, \varphi) \left( \sum_{r=1}^{N_{SNO}} (-\mu_r + n_r \cdot \ln \mu_r) - \frac{1}{2} \sum_{k=1}^{N_{SK}} \frac{[x_k - (aF\sin(\omega t_k + \varphi))]^2}{\sigma_{kp2}^2} \right) \quad (16),$$

where for the part of the expression pertaining to Super-Kamiokande the following conditions hold [8]

$\sigma_{kp2} = \sigma_{kdown}$ if $x_k > (AF\sin(\omega t_k + \varphi))$ and $\sigma_{kp2} = \sigma_{kup}$ if $x_k < (AF\sin(\omega t_k + \varphi))$.

and

$\sigma_{kp1} = \sigma_{kdown}$ if $x_k > 0$ and $\sigma_{kp1} = \sigma_{kup}$ if $x_k < 0$,

while for the part concerning SNO the relations (8) and (9) are valid. Also for the cumulative spectrum the Wilks' theorem ensures the exponential distribution for the amplitude of each individual frequency, under the null hypothesis.

The spectrum of the overall data as stemming from the relation (16), plotted up to the limit of 50 cycles/year so to take into account the coarse binning of the SK time series, is reported in Fig. 27. A very interesting result, which is a valuable systematic check of the applied method, is that the annual modulation line stand-ups very clearly, much more evidently with respect to the two experimental datasets taken separately. The frequency of the top of this line appears to be 0.94 cycles/year, thus close to the nominal unitary value.

The four highest peaks, excluding the annual line since we know that corresponds to a real signal, have, respectively, ordinates 8.10 (39.30), 7.75 (38.22), 6.53 (12.70) and 6.47 (18.77) (as usual in parenthesis there are the corresponding frequencies in cycles/year).

The null hypothesis Monte Carlo distributions are reported in Fig. 28, where they are plotted together with the model function (6) computed for the optimum M value equal to 437. The significances inferred from the Monte Carlo of the four peaks listed above are 15% for the highest, 2.21% for the second highest, 3.62% for the third highest peak, and 0.88% for the fourth highest peak. Thus, essentially the cumulative spectrum appears overall compatible with the constant rate hypothesis, especially taking into account the significance of the first peak (even though it should be noted the high significance featured by the fourth peak).

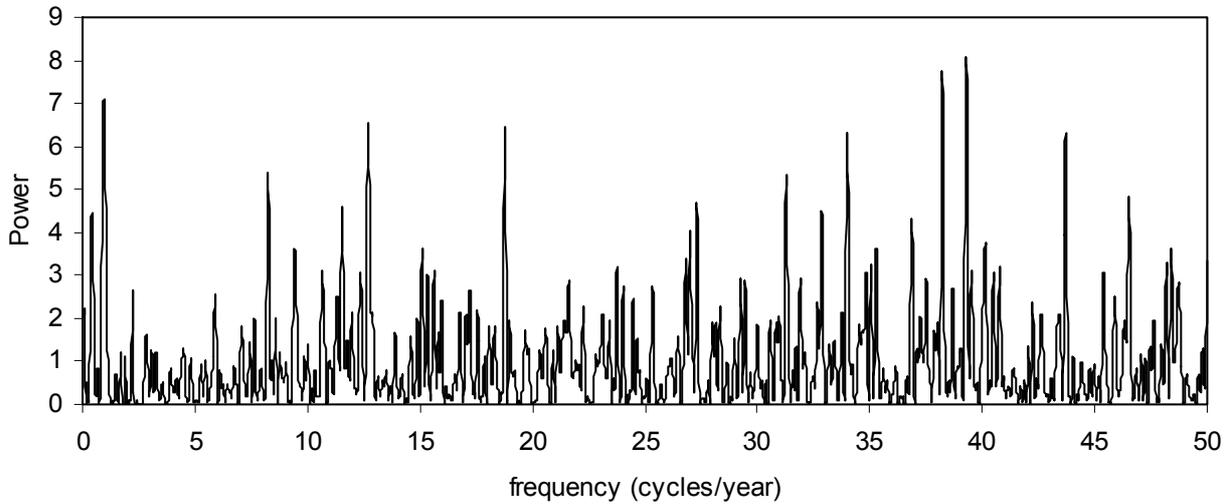

*Fig. 27 – Cumulative spectrum obtained from the entire SNO and SK datasets taken together.*

For completeness, it must be pointed out that comparing the main lines of this spectrum with those of the Super-Kamiokande I spectrum (Fig. 29 of [8]) it comes out that only one of the four major peaks of the latter is also among the four major peaks of the cumulative spectrum: this is the line at 39.30 cycles/year, which was the second highest in the SK spectrum, while it is the highest in the cumulative spectrum. On the other hand, the line at 9.42 cycles/line which is the major feature in the Super-Kamiokande spectrum is strongly suppressed in the cumulative spectrum; hence the joint, global analysis confirms in this respect the result reported in the above paragraph VI.

As said before, another interesting feature of the cumulative spectrum is the clear appearance of the annual modulation line, which is the third highest spectral line in the spanned frequency range. Neither in the Super-Kamiokande spectrum nor in the SNO spectrum such an expected line is so prominent. This peculiarity represents a good systematic check of the methodology chosen to write down the cumulative spectrum: with more statistics the manifestation in the spectrum of the annual modulation signal is indeed enhanced, as it should be expected.

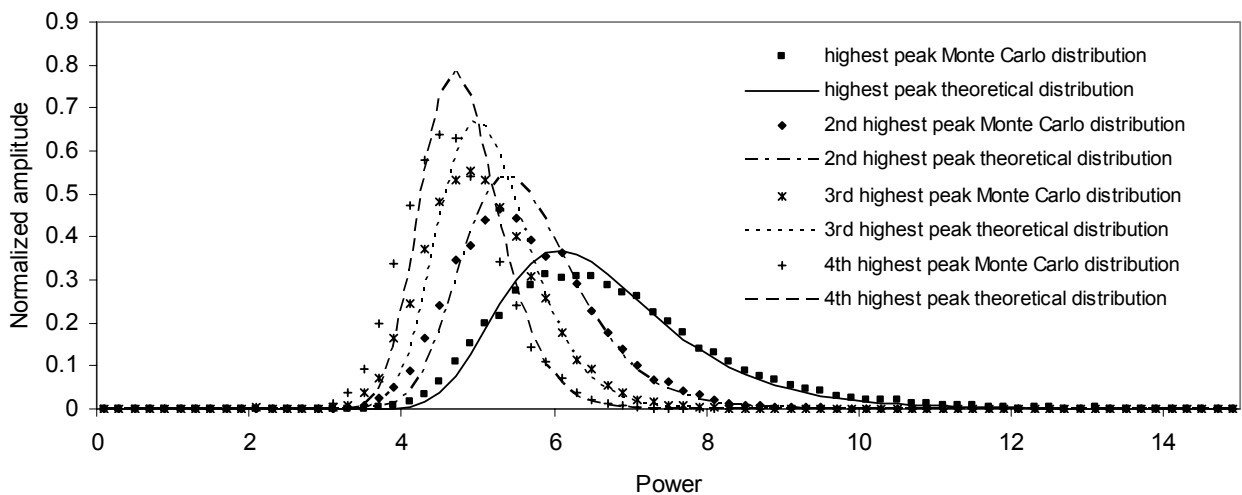

*Fig 28 – Cumulative spectrum null hypothesis distributions for the SK and SNO data taken altogether. The overlapped curves are the model functions (6) plotted with M= 437*

Such a circumstance makes it interesting to express in a quantitative way the significance of the identification of this line within the theoretical framework of the statistical detection of a signal of known (frequency) location. For this purpose we can simply compare the spectral value at the exact frequency of 1 cycles/year with the single frequency exponential noise distribution. Since the spectral value at that frequency is 5.83, the probability to get by chance a value as high or higher is then 0.0029: hence the confidence level of the detection of the line is 99.71%.

Such a significance obtained on the SNO-plus-Super-Kamiokande dataset is a remarkable improvement with respect to the significance of the detection of the annual line in the SNO data alone, which was quoted by the SNO collaboration in [1] at the level of 9.5% (i.e. a confidence level of 90.5%); this value represents also a further strong statistical confirmation of the solar origin of the data of the Cerenkov experiments, in addition to the association with the Sun direction used by the experimentalists to pick-up the true neutrino signals among the overwhelming detector backgrounds.

This result, whilst interesting per se, is especially significant also for its implications with respects to the future sub-MeV solar neutrino experiments based on the liquid scintillation technique, like Borexino and KamLAND solar phase [16][17]. Because of the lack of directionality in the scintillation light, these experiments will miss the demonstration of the solar origin of the events through the association with the Sun direction. The result obtained here, however, points out that with enough statistics and a good control of the systematic effects, as it appears to be the case for SNO and Super-Kamiokande, Borexino and KamLAND will be in condition to prove convincingly the solar origin of their data through the identification at high confidence level of the annual modulation signal.

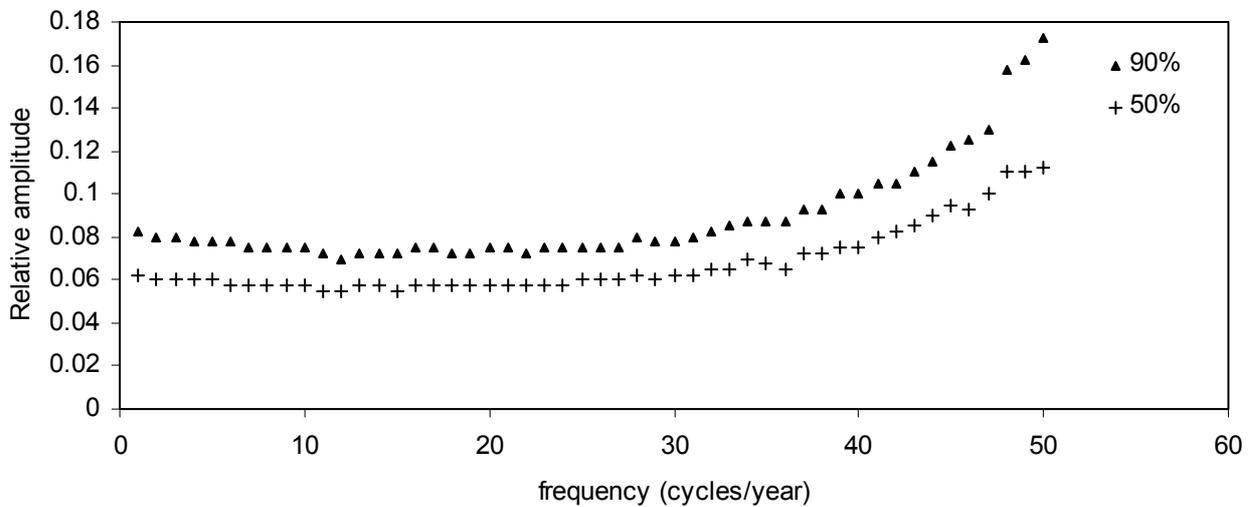

*Figure 29 – 50% and 90% sensitivity contours at the 99% confidence level for the cumulative analysis which takes into account all the available SNO and SK data.*

Finally, we repeated also in this case the sensitivity exercise, computing the iso-sensitivity contours for the combined dataset. The 50% and 90% contours, again for a false alarm probability amounting to 1%, are shown in Fig. 29. We note that up to about 43 cycles/year the sensitivity of the combined spectrum outperforms that of the spectrum of the SNO data taken alone: for example a low frequency 6% amplitude modulation would be discovered 50% of the times at the confidence level of 99%, whilst for SNO alone the same performances pertain to an 8% modulation amplitude.

Above 43 cycles/year the "tension" between the different binning of the two datasets degrades the performances of the combined spectrum with respect to the SNO data alone, even though they remain better than those of the Super-Kamiokande data alone (see Fig. 38 of [8]).

## X. CONCLUSIONS

The data released by the SNO collaboration related to the $D_2O$ and salt phases of the experiment have been searched for time modulations through the standard Lomb-Scargle method and a likelihood extension of the Lomb-Scargle method itself. Both methodologies, applied either on each dataset or on the combined dataset did not reveal any hint of periodicities, within the sensitivity limits dictated by the noise (Poisson scattering) affecting the data series. The sensitivity to true oscillations has been evaluated, as well, leading to a discovery probability (in case of the likelihood method) of 50% for a low frequency oscillation with relative amplitude of 8% at the confidence level of 99%. The data have been specifically examined for the periodicity of 9.42-9.43 cycles/year stemming from some analyses of the Super-Kamiokande I data, showing that the SNO results are strongly disfavouring an oscillation at that frequency in the 6-7% amplitude range in the $^8B$ solar neutrino flux. Finally, the joint study of the SNO and SK data, besides confirming the constancy of the neutrino flux over both experimental series, within the global experimental sensitivity which has been also evaluated, provides a 99.7% confidence level for the detection of the expected annual time variation signal, representing a further proof of the solar origin of the experimental datasets.

## AKNOWLEDGMENTS

The authors would like to thank the SNO Collaboration for making the data publicly available. G. Ranucci acknowledges also the very fruitful exchange of communications with Scott Oser.